\documentclass{ws-mpla}

\begin{document}

\markboth{Xin-Hui Zhang, Yu-Xiao Liu and Yi-Shi Duan}
{Localization of fermionic fields on braneworlds with bulk tachyon
matter}

\catchline{}{}{}{}{}

\title{LOCALIZATION OF FERMIONIC FIELDS ON BRANEWORLDS WITH BULK TACHYON MATTER}

\author{Xin-Hui Zhang,
 Yu-Xiao Liu\footnote{Corresponding author. E-mail: liuyx@lzu.edu.cn},
 Yi-Shi Duan}
\address{Institute of Theoretical Physics, Lanzhou University,
   Lanzhou 730000, P. R. China \\
   \vspace{2mm}
Email: zhangxingh03@lzu.cn, liuyx@lzu.edu.cn, ysduan@lzu.edu.cn }

\maketitle

\begin{abstract}

Recently, Pal and Skar in [arXiv:hep-th/0701266] proposed a
mechanism to arise the warped braneworld models from bulk tachyon
matter, which are endowed with a thin brane and a thick brane. In
this framework, we investigate localization of fermionic fields on
these branes. As in the 1/2 spin case, the field can be localized
on both the thin and thick branes with inclusion of scalar
background. In the 3/2 spin extension, the general supergravity
action coupled to chiral supermultiplets is considered to produce
the localization on both the branes as a result.

\keywords{Braneworlds;  Fermion zero modes.}

\end{abstract}

\ccode{PACS Nos.: 11.10.Kk,  04.50.+h.}

\section{introduction}

Extra dimensions were introduced to solve classical problems of
Particle Physics. 
In the 1920's, Kaluza and Klein \cite{KK1,KK2} proposed a theory
with a compact fifth dimension to unify electromagnetism with
Einstein gravity. Ref. \refcite{Antoniadis1990} contains the first
proposal for using large extra dimensions in the Standard Model
with gauge fields in the bulk and matter localized on the orbifold
fixed points (although the word brane was not used). In the course
of the last few years there has been some considerable activity in
the study of models that involve new extra spatial
dimensions.\cite{ADD,AADD,RS1,RS2}
Recent years have been witnessing a phenomenal interest in the
possibility that our observable four-dimensional (4D) Universe may
be viewed as a hypersurface (brane) embedded in a higher-dimensional
bulk space
with non-factorizable warped
geometry.\cite{braneworld1,braneworld2,braneworld3,braneworld4} In
this scenario, we are free from the moduli stabilization problem
in the sense that the internal manifold is noncompact and does not
need to be compactified to the Planck scale any more, which is one
of reasons why this new compactification scenario has attracted so
much attention.

Physical matter fields are confined to this hypersurface, while
gravity can propagate in the higher-dimensional space-time as well
as on the brane. A major issue in the so-called braneworld models
is the localization
problem\cite{localization1}$^{-}$\cite{Neupane2} of the Standard
Model fields on the brane. The most popular model in the context
of brane world theory is that proposed by Randall and Sundrum (RS
model).\cite{RS1,RS2} The localization mechanism has been widely
investigated in the RS model in five dimensions: Spin 0 field is
localized on a brane with positive tension.\cite{Spin0} Spin 1
field is not localized neither on a brane with positive tension
nor on a brane with negative tension.\cite{Spin1} Spin 1/2 and 3/2
fields are localized not on a brane with positive tension but on a
brane with negative tension.\cite{localization1} The general
observation is that the graviton, which is allowed to be free to
propagate in the bulk, are confined to the brane because of the
warped geometry, the massless scalar fields have normalizable zero
modes on branes of different types,\cite{Spin0} the Abelian vector
fields are not localized in the RS model in five dimensions but
can be localized in some higher-dimensional generalizations of
it,\cite{vector} however, the fermions do not have normalizable
zero modes both in five dimensions\cite{Spin0} and on a
string\cite{string1,string2} in six dimension.\cite{braneworld2}

In most of the models, the induced metric on the brane is scaled
Minkowski, i.e., the brane is flat. A few braneworld models
considered the curvature of the embedded brane, some of which
study the scenario with a FRW metric.\cite{FRW1,FRW2} What turns
out is that none of the braneworld models compatible to a FRW
metric on the brane could provide an warped geometry. However,
recently Ref. \refcite{skar} proposed an exact, warped braneworld
model arose from tachyon matter, falling in the class of exact
higher dimensional warped spacetimes. The bulk Einstein equations
can be exactly solved to obtain warped spacetimes. The solutions
thus derived are single brane models--one being a thin brane while
the other is of the thick variety. But they can not explain the
problem of fermionic localization. Why the spin 1/2 field can be
localized on the thick brane but not on the thin brane? What can
we do to have it localized on both the branes? And how about the
spin 3/2 field?

Based on this new braneworld scenario, we intend to investigate
the localization of the spin 1/2 and 3/2 fermionic fields. We know
that fermion interaction with a scalar domain wall can lead to
localization of chiral fermions\cite{addaction1} and a Yukawa-type
coupling to a scalar field of a domain-wall type can result in
chirality as well as localization of the
fermions.\cite{seif,wise1,wise2} It becomes necessary to allow
additional non-gravitational interaction to get spinor fields
confined to both the thin and the thick branes. The aim of the
present article is to introduce dynamics to determine the
localization of fermionic fields. We shall prove that spin 1/2 and
3/2 fields can be localized on both the branes.

\section{The review of braneworld models arising from tachyon matter}

In what follows, let us start with a brief review of the new
braneworld models proposed by Ref. \refcite{skar}. Consider the
action in five dimensional spacetime
\begin{equation}\label{action}
S=\int{d^5x\sqrt{-g}\,\left(\frac{1}{2\kappa_5^2}(R-2\Lambda_5) +
V(T) \sqrt{1+g^{MN}\partial_MT\partial_NT} \right)
}+\lambda_b\int{d^4 x\sqrt{-\hat{g}}},
\end{equation}
where  $\Lambda_5$, $\lambda_b$ and $V(T)$ are the bulk
cosmological constant, the brane tension and the potential of the
tachyon field $T$, respectively,
$\hat{g}=\det{(\hat{g}_{\mu\nu})}$ with $\hat{g}_{\mu\nu}$ being
the induced metric on the brane. The first term in action
(\ref{action}) is the contribution from the pure 5D gravity and
the tachyon field, and the second one is that from the brane. The
FRW metric of the bulk spacetime is taken to be of the warped form
\begin{eqnarray}
ds^2&=& e^{2f(\sigma)}\hat{g}_{\mu\nu}(x)
dx^{\mu}dx^{\nu}+d\sigma^2
\nonumber \\
 &=&e^{2f(\sigma)}[-dt^2+a^2(t)(dx^2+dy^2+dz^2)]+d\sigma^2,
\end{eqnarray}
where $a(t)$ is the scale factor and $f(\sigma)$ is the warp factor.
Since the model is restrict to a de-Sitter brane, the scale factor
can be expressed as $a(t)=e^{Ht}$ with $H$ being the Hubble
constant. Performing a conformal transformation $d\tau=a^{-1}(t)dt$,
one can rewrite the bulk metric as
\begin{equation}\label{metric}
ds^2=e^{2f(\sigma)}a^2(\tau)[-d\tau^2+dx^2+dy^2+dz^2]+d\sigma^2.\textbf{}
\end{equation}

For simplicity, the tachyon field can be chosen to have the form
$T=T(t,\sigma)$. From the form of the Einstein tensors:
$G_{MN}=-\Lambda_5g_{MN}+\kappa^2_5T_{MN}$, since there is no
off-diagonal term, one could have either the time-derivative or the
$\sigma$-derivative of the tachyonic field vanish. In this context,
set the time-derivative of $T$ to be zero, and choose the tachyon
field $T$ to be the form of $T(\sigma)$. Consequently, Einstein
equations reduce to the following two coupled differential equations
\begin{eqnarray}
3H^2e^{-2f}-6f'^2-3f''&=&\Lambda_5-\kappa_5^2V(T)\label{equation1}
\sqrt{1+T'^2}-\kappa_5^2\lambda_b\delta{(\sigma-\sigma_0)},\\
-6H^2e^{-2f}+6f'^2&=&-\Lambda_5+\kappa_5^2\frac{{V(T)}}{\sqrt{1+T'^2}}\label{equation2},
\end{eqnarray}
in which $\sigma_0$ is the definite location of the brane. Given an
expression for the tachyon potential $V(T)$, one can obtain a set of
solutions for the above equations.

\subsection{The thin brane}

A thin brane is realized as a sharp peak of the warp factor at a
definite location $\sigma_0$ in the entire range of the extra
dimension. To get a thin brane, the contribution of the brane
tension $\lambda_b$ must be considered in the action. It is not
required here to obtain the bulk geometry, so set $\Lambda_5=0$.
Taking the tachyon potential as the form of
\begin{equation}
V(T)=\frac{3\sqrt{2}\kappa^2}{\kappa^2_5}\sin(\sqrt{2}\kappa|T|)
[2-\cos^2(\sqrt{2}\kappa|T|)]^{\frac{1}{2}}
\end{equation}
and solving Eqs. (\ref{equation1}) and (\ref{equation2}), one can
obtain the following expressions for the warp factor and the tachyon
field
\begin{eqnarray}
f(\sigma)&=&-\kappa|\sigma|,\\
T(\sigma)&=&\frac{1}{\sqrt{2}\kappa}\cos\left(\frac{\kappa}{H}e^{\kappa|\sigma|}\right).
\end{eqnarray}
One can see that the warp factor has discontinuous derivative at the
brane location.

\subsection{The thick brane}

For a thick brane, one need not consider the contribution from the
brane tension. Hence, in this case, the intention is to solve the
Einstein equations (\ref{equation1}) and (\ref{equation2}), with
$\Lambda_5\neq{0}$ but $\lambda_b=0$. The tachyon potential,
expressed as a function of $\sigma$, is given by
\begin{eqnarray}
V(\sigma)&=&\frac{1}{\kappa^2_5}\texttt{sech}^2(b\sigma)\sqrt{(\Lambda_5+6b^2)\sinh^2(b\sigma)
+\Lambda_5+6(b^2-H^2)}\times\\\nonumber
&&\sqrt{(\Lambda_5+6b^2)\sinh^2(b\sigma)+(\Lambda_5-6H^2)},
\end{eqnarray}
where $b$ is a certain arbitrary constant. Then, the exact
expressions of $f(\sigma)$ and $T(\sigma)$ are solved to be
\begin{eqnarray}
f(\sigma)&=&\ln{\cosh(b\sigma)}, \\
T(\sigma)&=&-\frac{i}{b}\sqrt{\frac{3(b^2+H^2)}{\Lambda_5-6H^2}}
\textrm{EllipticF}\left[ib\sigma,
\frac{\Lambda_5+6b^2}{\Lambda_5-6H^2}\right].
\end{eqnarray}
These special solutions would be utilized to analyse localization
of fermionic fields on these branes in the next section.

\section{Localization of Spin 1/2 Fermions}

In this section we investigate  localization of a spin 1/2 fermionic
field. Let us consider the Dirac action of a massless spin 1/2
fermion in five dimension
\begin{equation}\label{action-Dirac}
S_{1/2}=\int{d}^5x\sqrt{-g}\bar{\Psi}\Gamma^MD_M\Psi,
\end{equation}
from which the Dirac equation is given by
\begin{equation}\label{equation-Dirac}
\gamma^{\bar{M}} E^M_{\bar{M}}(\partial_M+\omega_M-ieA_M)\Psi=0,
\end{equation}
where the indices of 5D coordinates are labeled with Latin letters
$M, N,...$ while $\bar{M},\bar{N},...$ denote the local lorentz
indices. $\Gamma^M$ and $\gamma^{\bar{M}}$ are the curved gamma
matrices and the flat ones, respectively, which are connected by the
relation $\Gamma^M=E^M_{\bar{M}}\gamma^{\bar{M}}$ with
$E^M_{\bar{M}}$ being the vielbein,
$\omega_M=\frac{1}{4}\omega^{\bar{M}\bar{N}}_M\gamma_{\bar{M}}\gamma_{\bar{N}}$
is the spin connection, and $A_M$ is a U(1) gauge field. In this
model, we consider the special case with $A_M=0$. The spin
connection $\omega^{\bar{M}\bar{N}}_M$ in the covariant derivative
$D_M\Psi=(\partial_M+\frac{1}{4}\omega^{\bar{M}\bar{N}}_M\gamma_{\bar{M}}\gamma_{\bar{N}})\Psi$
is defined as
\begin{eqnarray}\label{vielbein}
\omega^{\bar{M}\bar{N}}_M&=&\frac{1}{2}E^{N\bar{M}}(\partial_ME^{\bar{N}}_N-\partial_NE^{\bar{N}}_M)
 \nonumber\\
&-&\frac{1}{2}E^{N\bar{N}}(\partial_ME^{\bar{M}}_N-\partial_NE^{\bar{M}}_M)\nonumber \\
&-&\frac{1}{2}E^{P\bar{M}}E^{Q\bar{N}}(\partial_PE_{Q\bar{R}}-\partial_QE_{P\bar{R}})E^{\bar{R}}_M.
\end{eqnarray}
In terms of Eqs. (\ref{metric}) and (\ref{vielbein}), the
non-vanishing components of $\omega_M$ are
\begin{equation}
\omega_\mu =
\frac{1}{2}f'(\sigma)\Gamma_\mu\Gamma_\sigma+\hat{\omega}_\mu,
\end{equation}
where $\mu=0,1,2,3$ and $\hat{\omega}_\mu=\frac{1}{4}
\bar\omega_\mu^{\bar{\mu} \bar{\nu}} \Gamma_{\bar{\mu}}
\Gamma_{\bar{\nu}}$ is the spin connection derived from the metric
$\hat{g}_{\mu\nu}(x)=\hat{e}_{\mu}^{\bar{\mu}}
\hat{e}_{\nu}^{\bar{\nu}}\eta_{\bar{\mu}\bar{\nu}}$. The Dirac
equation (\ref{equation-Dirac}) becomes
\begin{equation}
\left\{\Gamma^{\mu}\left(\hat{D}_\mu+
\frac{1}{2}f'(\sigma)\Gamma_\mu\Gamma_\sigma\right)
+\Gamma^\sigma\partial_\sigma\right\}\Psi=0,
\end{equation}
where $\hat{D}_\mu=\partial_\mu+\hat{\omega}_\mu$ is the Dirac
operator on the brane.

 Now, considering the conditions $\Gamma^\sigma=\gamma^\sigma$
and $\gamma^\sigma\Psi=\Psi$, from the above equation, we can obtain
the solutions of the form $\Psi(x^M)=\psi(x^\mu)U(\sigma)$, where
$U(\sigma)$ satisfies
\begin{eqnarray}
\left(\partial_\sigma+2f'(\sigma)\right){U(\sigma)}&=&0,
\end{eqnarray}
i.e.
\begin{equation}
U(\sigma)=U_0 \, e^{-2f(\sigma)}.
\end{equation}
Substituting the solution back into the 5D Dirac equation in
curved space, Ref. \refcite{skar} has shown that the integral of
the Dirac action (\ref{action-Dirac}) diverges for the thin brane
model whereas it is finite for the thick brane model. Now let us
include a real scalar field $\phi$ in order to solve this problem.
The modification of the Dirac action will be through some Yukawa
term, with the coupling $\lambda$
\begin{equation}
S_{1/2}=\int{d}^5x\sqrt{-g}[\bar{\Psi}\Gamma^M (\partial_M+\omega_M
)\Psi+\lambda\bar\Psi\phi\Psi], \label{action-Dirac2}
\end{equation}
and the corresponding equation of motion is
\begin{equation}
\left\{\Gamma^{\mu}\left(\hat{D}_\mu+
\frac{1}{2}f'(\sigma)\Gamma_\mu\Gamma_\sigma \right)+
\Gamma^\sigma\partial_\sigma +\lambda\phi(\sigma) \right\}\Psi=0.
\end{equation}
The detail of the $\phi$-field dynamics will not be important for
our discussion. Imposing the relation $\Gamma^\sigma=\gamma^\sigma$
and the chirality condition $\gamma^\sigma\Psi=+\Psi$, we only need
to solve
\begin{eqnarray}
\left(\partial_\sigma+2f'(\sigma)+
\lambda\phi(\sigma)\right){U(\sigma)}&=&0.
\end{eqnarray}
The solution of the above equation is turned out to be
\begin{equation}
\Psi(x^M)=U_0e^{-2f(\sigma)-\lambda\int^\sigma\phi(\sigma)d\sigma}\psi(x^\mu).
\end{equation}
In terms of this new variable, the action (\ref{action-Dirac2}) can
be rewritten as
\begin{equation}\label{action 1/2}
S_{1/2}=U_0^2\int_0^{\infty} d\sigma~
e^{-f(\sigma)-2\lambda\int^\sigma\phi(\sigma)d\sigma} \int{ d^4x
\sqrt{-\hat{g}}\bar\psi\gamma^\mu(\partial_\mu+\hat{\omega}_\mu)\psi}.
\end{equation}
In order to localize spin 1/2 fermion in this framework, the
integral (\ref{action 1/2}) should be finite. In fact, the
requirement is easily satisfied. Now let us look for the condition
for localization of spin 1/2 field. For example, we shall only
assume that the $\phi$ field equation of motion admits a localized
$\sigma$-dependent solution such that
$\phi(\sigma)\rightarrow|\upsilon|\epsilon(\sigma)$ as
$|\sigma|\rightarrow{\infty}$, where
$\upsilon=\langle\phi\rangle$, and $\epsilon(\sigma)$ is the sign
function i.e.
\begin{equation}\label{condition1}
\phi(\sigma)=|\upsilon|\epsilon(\sigma).
\end{equation}
Once again the second integral leads to the Dirac equation in curved
spacetime on the brane. So, a finite value for the first integral
involving $\sigma$ will guarantee that the zero mode of a spin 1/2
field is localized on the branes. One can readily show that for
large enough values of $\lambda|\upsilon|$, this integral
(\ref{action 1/2}) is finite for both the thin brane and the thick
brane. For the thin brane, it is sufficient to have
$\lambda|\upsilon|>\frac{\kappa}{2}$. We can choose the form as
follows
\begin{equation}\label{condition2}
\phi(\sigma)=c\sigma^{n},
\end{equation}
where $c>0$ is the arbitrary constant and $n$ satisfies $n\geq1$. We
can also choose the form
\begin{equation}\label{condition3}
\phi=e^{c\sigma},
\end{equation}
where $c$ is the arbitrary positive constant. So spin 1/2 field is
localized on both the branes under condition (\ref{condition1}) or
(\ref{condition2}) or (\ref{condition3}). Of course, there are many
other choices which result in a finite action.

\section{Localization of Spin 3/2 Fermions}

Next we turn to a spin 3/2 field, i.e. the gravitino. Let us start
by considering the general supergravity action coupled to chiral
supermultiplets\cite{action3/2}
\begin{eqnarray}
S_{sg}&=&\int{d}^5x\sqrt{-g} \frac{1}{2\kappa_5^2}R
 -  \int{d}^5x\sqrt{-g} G^i_jg^{MN}D_M\phi_iD_N\phi^{*j} \nonumber \\
 &&+ \int{d}^5x\sqrt{-g}\bar\Psi_M{i}\Gamma^{[M}\Gamma^N
                \Gamma^{R]}\{D_N\Psi_R
    +i\lambda\Gamma^\sigma(G^iD_N\phi_i-G_iD_N\phi^{*i})\Psi_R\}
 +...,
\label{action3/2}
\end{eqnarray}
where $D_N\phi_i=\delta^\sigma_N\partial_\sigma\phi_i$, $\lambda$
denotes the coupling constant, the indices i, j represent the number
of the scalar fields, and the square bracket on the curved gamma
matrices $\Gamma^M$ denotes the anti-symmetrization with weight 1.
The function $G$ is the Kahler potential
$G(\phi^*,\phi)=-3\log(-\frac{K(\phi^*,\phi)}{3})+\log|W(\phi)|^2$,
with a general function $K$ and the superpotential $W$. Without loss
of generality, the expression $K=-3\exp(-\phi^*\phi/3)$ can be
chosen. Moreover, various derivations of the Kahler potential are
defined as
\begin{eqnarray}
G^i=\frac{\partial{G}}{\partial\phi_i},\;\;
G_i=\frac{\partial{G}}{\partial\phi^{*i}}\;\;
G^i_j=\frac{\partial^2G}{\partial\phi_i\partial\phi^{*j}}.
\end{eqnarray}
The equations of motion for the Rarita-Schwinger gravitino field are
derived as
\begin{equation}\label{equation-grav}
\Gamma^{[M}\Gamma^N
\Gamma^{R]}(D_N+i\lambda\Gamma^\sigma\delta^\sigma_N\phi^*\overleftrightarrow{\partial_\sigma}\phi)\Psi_R=0.
\end{equation}
Here $D_N\Psi_R=\partial_N\Psi_R-\Gamma^M_{NR}\Psi_M+\omega_N\Psi_R$
and $\phi^*\overleftrightarrow{\partial_\sigma}\phi=
\phi^*\partial_\sigma\phi-\partial_\sigma\phi^*\phi$, which is
defined with the affine connection
$\Gamma^R_{MN}=E^R_{\bar{M}}(\partial_ME^{\bar{M}}_N+\omega^
{\bar{M}\bar{N}}_ME_{N\bar{N}})$. With the non-vanishing components
of spin connection and affine connection
\begin{eqnarray}
\omega_\mu &=&
\frac{1}{2}f'(\sigma)\Gamma_\mu\Gamma_\sigma+\hat{\omega}_\mu,
\\\nonumber
\Gamma^\nu_{\mu\sigma}&=&f'(\sigma)\delta^\nu_\mu,\\\nonumber
\Gamma^\sigma_{\mu\nu}&=& -a^2(\tau)e^{2f(\sigma)}\eta_{\mu\nu},
\end{eqnarray}
the components of the covariant derivative are calculated as follows
\begin{eqnarray}
D_\mu\Psi_\nu&=&\partial_\mu\Psi_\nu+a^2(\tau)e^{2f(\sigma)}\eta_{\mu\nu}\Psi_\sigma+
\frac{1}{2}f'(\sigma)\Gamma_\mu\Gamma_\sigma\Psi_\nu+\hat{\omega}_\mu\Psi_\nu,\\
D_\mu\Psi_\sigma&=&\partial_\mu\Psi_\sigma-f'(\sigma)\delta^\nu_\mu\Psi_\nu+
\frac{1}{2}f'(\sigma)\Gamma_\mu\Gamma_\sigma\Psi_\sigma+\hat{\omega}_\mu\Psi_\sigma,\\
D_\sigma\Psi_\mu&=&\partial_\sigma\Psi_\mu-f'(\sigma)\delta^\nu_\mu\Psi_\nu.
\end{eqnarray}
According to the assumption $\Psi_\sigma=0$, we will look for the
solutions of the form
\begin{equation}
\Psi_\mu(x,\sigma)=\psi_\mu(x){U(\sigma)},
\end{equation}
where $\psi_\mu(x)$ satisfies the following equations
$\gamma^\nu\psi_\nu=\partial^\mu\psi_\mu=\gamma^
{[\mu}\gamma^\nu\gamma^{\rho]}\hat{D}_\nu\psi_\rho=0$ and
$\Gamma^\sigma\psi_\mu=\psi_\mu$. Then the equations of motion
(\ref{equation-grav}) reduce to
\begin{equation}
(\partial_\sigma+f'(\sigma)+i\lambda\phi^*\overleftrightarrow{\partial_\sigma}\phi)U(\sigma)=0,
\end{equation}
from which $U(\sigma)$ is easily solved to be
\begin{equation}\label{zero-grav}
U(\sigma)=U_0e^{-f(\sigma)-i\lambda\int^\sigma\phi^*\overleftrightarrow{\partial_\sigma}\phi{d}\sigma}.
\end{equation}
To this end, let us substitute the zero mode (\ref{zero-grav})
into the action (\ref{action3/2}) and consider that
$i\lambda\phi^*\overleftrightarrow{\partial_\sigma}\phi$ is real.
It turns out that the action of the Rarita-Schwinger gravitino
field becomes
\begin{eqnarray}\label{action3/22}
S_{3/2} &=& \int {d^5}x \sqrt{-g} i\bar\Psi_M
    \Gamma^{[M}\Gamma^N \Gamma^{R]}(D_N
   +i\lambda\Gamma^\sigma\delta^\sigma_N
   \phi^*\overleftrightarrow{\partial_\sigma}\phi)\Psi_R \nonumber
   \\
 &=&U_0^2\int_0^\infty{d\sigma}{e}^{-f(\sigma)-2i\lambda\int^\sigma\phi^*
\overleftrightarrow{\partial_\sigma}\phi{d\sigma}} \int {d^4}x
\sqrt{-\hat{g}} i\bar\psi_\mu
\gamma^{[\mu}\gamma^\nu\gamma^{\rho]} \hat{D}_\nu \psi_\rho.
\end{eqnarray}
A finite value for the first integral will guarantee that the zero
mode of a spin 3/2 field is localized on the brane. The requirement
is easily satisfied. We shall
 assume that the $\phi$ field satisfies
\begin{equation}\label{condition11}
\phi=|\upsilon|e^{i|\sigma|},
\end{equation}
where $\upsilon$ is a constant, then
\begin{equation}
i\lambda\phi^*\overleftrightarrow{\partial_\sigma}\phi
=-2\lambda|\upsilon|^2\epsilon(\sigma).
\end{equation}
 When
$\lambda<-\frac{\kappa}{4|\upsilon|^2}$, the integral
(\ref{action3/22}) is finite for the thin brane model with
decreasing warp factor $f(\sigma)=-\kappa|\sigma|$, and it is also
finite for the thick brane model with increasing warp factor
$f(\sigma)=\ln{\cosh(b\sigma)}$. Thus, it turns out that the zero
mode of the gravtino field can be localized on both the thick brane
and the thin brane.

\section{Conclusion}

Since spin half fields can not be localized on both the branes by
gravitational interaction only, it becomes necessary to introduce
additional non-gravitational interaction to get spinor field
confined to both the branes. In the braneworld  model arising from
tachyon matter, we introduce dynamics which can determine the
location of the branes in the bulk.

In the spin 1/2 field, we introduce the scalar $\phi$ by involving
the Yukawa coupling. We obtain the general form (\ref{action
1/2}), which generalized the conclusion drew by \refcite{seif}. By
choosing the appropriate values of the scalar field $\phi$ and the
coupling constant $\lambda$, the action (\ref{action 1/2}) can be
finite on both the branes. That is to say, the spin 1/2 field can
be localized not only on the thick brane but also on the thin
brane. It has been proved that many choices can make the action
finite. In this paper, we only give the conditions
(\ref{condition1}), (\ref{condition2}), (\ref{condition3}) as
examples.

For the spin 3/2 field, i.e., the Rarita-Schwinger gravitino
field, we consider the supergravity action coupled to the chiral
supermultiplets and obtain the general form of the action
(\ref{action3/22}). It is proved that under the condition
(\ref{condition11}), the action can be finite, i.e.,  the spin 3/2
fields also can be localized on both the thin brane and the thick
brane.

Moreover, to localize the fermions on the brane or the string-like
defect, there are some other backgrounds could be considered
besides scalar field and gravity, for example, gauge
field\cite{LiuJHEP2007,ZhaoMPLA2007} or
vortex\cite{WangMPLA2005,DuanMPLA2006,LiuIJMPA2007} and gravity
backgrounds. The localization of the topological Abelian Higgs
vortex coupled to fermion can be fund in our another
work.\cite{LiuCTP2007}

\section*{Acknowledgments}

This work was supported by the National Natural Science Foundation
of the People's Republic of China (No. 502-041016, No. 10475034 and
No. 10705013) and the Fundamental Research Fund for Physics and
Mathematics of Lanzhou University (No. Lzu07002).

\end{document}